# Monte Carlo simulation of SARS-CoV-2 radiation-induced inactivation for vaccine development

Ziad Francis[1], Sebastien Incerti[2], Sara A. Zein[2], Nathanael Lampe[2,#], Carlos A. Guzman[3] and Marco Durante[4,5,*]

[1] Saint Joseph University, U.R. Mathématiques et Modélisation, Beirut, Lebanon

[2] Université de Bordeaux, CNRS/IN2P3, UMR5797, Centre d'Études Nucléaires de Bordeaux Gradignan, France

[3] Helmholtz Zentrum für Infektionsforschung (HZI), Department of Vaccinology and Applied Microbiology, Braunschweig, Germany

[4] GSI Helmholtzzentrum für Schwerionenforschung, Biophysics Department, Darmstadt, Germany

[5] Technische Universität Darmstadt, Institute of Condensed Matter Physics, Darmstadt, Germany

[#] this author now works for a private company

* Corresponding author: Marco Durante

**Email:** M.Durante@gsi.de

## Abstract

Immunization with an inactivated virus is one of the strategies currently being tested towards developing a SARS-CoV-2 vaccine. One of the methods used to inactivate viruses is exposure to high doses of ionizing radiation to damage their nucleic acids. Although $\gamma$-rays effectively induce lesions in the RNA, envelope proteins are also highly damaged in the process. This in turn may alter their antigenic properties, affecting their capacity to induce an adaptive immune response able to confer effective protection. Here, we modelled the impact of sparsely and densely ionizing radiation on SARS-CoV-2 using the Monte Carlo toolkit Geant4-DNA. With a realistic 3D target virus model, we calculated the expected number of lesions in the spike and membrane proteins, as well as in the viral RNA. We show that $\gamma$-rays produce significant spike protein damage, but densely ionizing charged particles induce less membrane damage for the same level of RNA lesions, because a single ion traversal through the nuclear envelope is sufficient to inactivate the virus. We propose that accelerated charged particles produce inactivated viruses with little structural damage to envelope proteins, thereby representing a new and effective tool for developing vaccines against SARS-CoV-2 and other enveloped viruses.

## Introduction

The threat from the new SARS-CoV-2 is omnipresent in all regions of the world *(1)*. Several companies and research groups are reporting premising rssults in the development of vaccines *(2–4)*. Some of the SARS-CoV vaccines developed so far follow the traditional approach based on inactivated or live attenuated viruses *(5–8)*. SARS-CoV-2 virus is sensitive to the standard inactivation methods used for producing vaccines against other viral diseases *(9)*. However, traditional chemical inactivation methods can lead to the modification of critical epitopes in both inactivated and subunit vaccines, thereby affecting overall vaccine efficacy *(10)*. Although live attenuated virus-based vaccines generally induce potent cellular immune response, antibody production is often less efficient, and the use of attenuated viruses is often not indicated in individuals who are immunosuppressed as a result of disease or therapeutic interventions. In the current pandemics, mRNA vaccines are emerging as very attractive alternatives, due to their ease of manufacture and potent immune response *(11)*, but their long-term effectiveness and safety remains to be elucidated in comparison to the well-established inactivated virus standard.

Among inactivation methods, radiation technology remains of interest to vaccine manufacturers, because it penetrates pathogens to damage the nucleic acid without residual chemical contaminants *(12,13)*. Nevertheless, the development of irradiated vaccines has not been pursued actively over the past 20 years for two main reasons. First, the development of new radiation techniques has been considered impractical or difficult due to issues accessing radiation equipment *(14)*. Second, it has been thought that modern subunit vaccines would provide a solution, as they can be developed more easily *(15)*. However, recent successful development of irradiated vaccines for human malaria *(16)* and influenza *(17,18)* have not only demonstrated the feasibility and practicality of this technique, but also the efficacy of such vaccination approaches, prompting research in applying the method to other viruses *(19,20)*. Yet it is known that high-dose $\gamma$-ray exposure can damage epitopes necessary to elicit the immune response *(21)*. Formaldehyde and other chemicals commonly used for virus inactivation also induce severe damage to the protein structures involved in the elicitation of adaptive immune response post vaccination *(22)*.

With conventional γ-ray sterilization, many photons are needed to induce lesions in RNA. However, the track structure of charged particles, such as those used in cancer therapy *(23)*, suggest that a single traversal is sufficient to induce lethal lesions in RNA with minimal membrane damage *(13)*. In this work we performed a comprehensive simulation of ionizing radiation damage to the SARS-CoV-2 using Geant4-DNA *(24)*, a Monte Carlo simulation toolkit which is used for modeling biological damage induced by ionizing radiation at the DNA scale. We developed a Geant4-DNA application to simulate the SARS-CoV-2 structure and calculated the damage to the virus' spike and membrane proteins, and the viral RNA following exposure to radiation of different quality, that is, low- or high- linear energy transfer (LET) radiation. High-LET (densely ionizing) radiation is in fact densely ionizing and may inactivate the virus with less membrane damage than low-LET (sparsely ionizing) radiation.

## Materials and Methods

### *Physics processes and irradiation particles*

The incident radiation is a parallel beam with a circular cross-section 132 nm in diameter, uniformly covering the complete geometry of the virus including the spike proteins. Simulations were carried out for 3 MeV α-particles, 450 MeV/amu $^{56}$Fe- ions, 80 MeV/amu $^{12}$C-ions, 2 keV and 200 keV electrons, 80 MeV and 30 MeV protons and a $^{60}$Co γ-ray spectrum (1.17 MeV and 1.32 MeV).

The Geant4-DNA processes *(25–28)* were used to simulate step-by-step tracks of ions and electrons. The Livermore models *(29,30)* were used to simulate gamma interactions. Processes included for protons and alpha particles were nuclear scattering, electronic excitation, ionization, and charge transfer. For electrons, elastic scattering, electronic excitation, ionization, vibrational and rotational excitations, and electron attachment were considered. Electrons were followed down to 7.5 eV, below this energy an electron goes one additional solvation step and its remaining energy is locally deposited. Heavy ions interactions consisted only of ionization which was simulated using the relativistic Rudd model with an effective charge scaling as described in *(31)* and previously used for carbon ion fragments *(32)*.

Since all the included processes of Geant4-DNA were validated for liquid water, we proceeded with a density scaling to account for the different virus materials. Therefore, the different protein densities were calculated based on the constitutions mentioned in *(33)* and the cross sections were scaled accordingly. The calculated densities are 1.4 g/cm$^3$ for the spike proteins, 1.38 g/cm$^3$ for the membrane protein, and 1.46 g/cm$^3$ for the inner sphere, equal to the RNA density.

## *Simulation of radiation-induced damage*

Energy deposition points were calculated in the spike proteins, membrane proteins, and RNA molecule.

The RNA size was calculated based on the genome sequence published by the National Center for Biotechnology Information *(34)*. The viral ss-RNA sequence length is 29903 bases and its mass is ~9,584 MDa. Considering the RNA density, we can estimate its approximate volume to be ~10900 nm$^3$. This value is equal to ~4% of the inner volume of the virus, the 80 nm sphere shown in Figure 1. Therefore, for each energy deposition point inside the virus a random sampling was applied with a 4% probability that this point is positioned on the RNA. A recent study *(35)* using the Monte Carlo code TOPAS investigated different RNA setups with fast ions irradiation showing differences in the extreme case where RNA is densely centered in the middle of the capsid. The assumption that RNA is homogeneously distributed inside the virus, was previously used in *(36)* for DNA damage assessment after proton irradiations, leading to reasonable results. While modelling the RNA at the atomistic level is still the reference method, the approximation of a homogeneous geometrical cross section is simpler to incorporate in such comparative studies. In an earlier study *(36)* this method was validated for DNA irradiation where the obtained single and double strand break ratios were in agreement with experiments and detailed atomistic calculations

*(37)*. Although this approach was never validated for viral RNA, due to lack of specific RNA irradiation data, we could consider that RNA geometry and positioning vary within a population of viral structures. Therefore, considering a large amount of virus targets, with different RNA distributions, irradiated by a uniform beam, the homogeneous RNA cross section is an average approximation. Moreover, other simulations carried out while changing the cross section relative value by 25%, meaning considering 3% RNA volume instead of 4% does not affect our results by a visible amount, since our conclusions rely mainly on comparisons among different particle types. Although the uniform RNA model presented here is suitable for numerical simulations, it doesn't provide a differentiation between sugar-phosphate and nitrogenous bases damage. Therefore, any RNA damage was considered sufficient to inactivate the virus.

Studies from the literature on DNA damage reported breaks induced by low energy electrons. The incident energies ranged between 1 eV and 30 eV *(38–41)*, and damage was induced by energies as low as 3-5 eV. Taking these values into account we determined that an energy deposition above a threshold of 10 eV, positioned on the RNA volume, is considered a damage.

For spike proteins, if the sum of energy depositions inside one protein was higher than the 10 eV damage energy threshold, we consider the protein as damaged. The number of damaged proteins was counted and normalized to a single RNA damage. As we are looking for the ionizing radiation type that would yield the lowest amount of protein damage per RNA damage to inactivate the virus, this normalization is desirable. In a second stage, in order to verify the influence of the damage threshold on our results, calculations were evaluated for 4 different threshold values, 5 eV, 10 eV, 15 eV and 20 eV.

# Results

### *Simulation setup*

The Geant4-DNA extension *(24,26–28)* of the Geant4 Monte-Carlo toolkit *(30,42,43)* was used to simulate ionizing particle tracks and energy deposition inside the virus model. The virus membrane was represented by a spherical volume, the outer and inner diameters were set to 100 nm and 80 nm, respectively. These dimensions are based on experimental data from the literature. The virus size reported by published experiments varies between 60 nm and 140 nm *(44–47)*, meaning an average of 100 nm is a reasonable estimation. 166 membrane proteins were modelled, represented by cylindrical volumes 6 nm in diameter and 6 nm in height homogeneously distributed within the 10 nm thick membrane. The spike proteins were modeled by conical volumes with base and top diameters of 6 nm and 14 nm, and 16 nm height. A total of 141 spike protein volumes were homogeneously distributed on the outside of the membrane. Figure 1 shows the configuration of the described virus structure. The approximate dimensions of the proteins of the SARS-CoV-2 were estimated based on the descriptions from the protein data bank *(48,49)*. The nucleic acid contained in the envelope is a ss-RNA of approximately 30 kbp *(1)*.

### *Damage distribution*

For a comparison of sparsely and densely ionizing radiation in the SARS-CoV-2 target, we show in Figure 2 the simulation of the damage produced in the virus by: $^{60}$Co $\gamma$–rays (LET=0.24 keV/$\mu$m), the standard radiation used for virus inactivation in vaccine development; 3 MeV $\alpha$-particles, a common standard in high-LET radiation (LET=126 keV/$\mu$m) with a range of only 18 μm in water;

and $^{56}$Fe 450 MeV/amu (LET=178 keV/μm), a densely ionizing ion often used at accelerators in studies of space radiation protection *(50)* and with a range of approximately 12 cm in water. Probability distributions of the number of RNA damage sites were normalized per 1 eV of energy deposition.

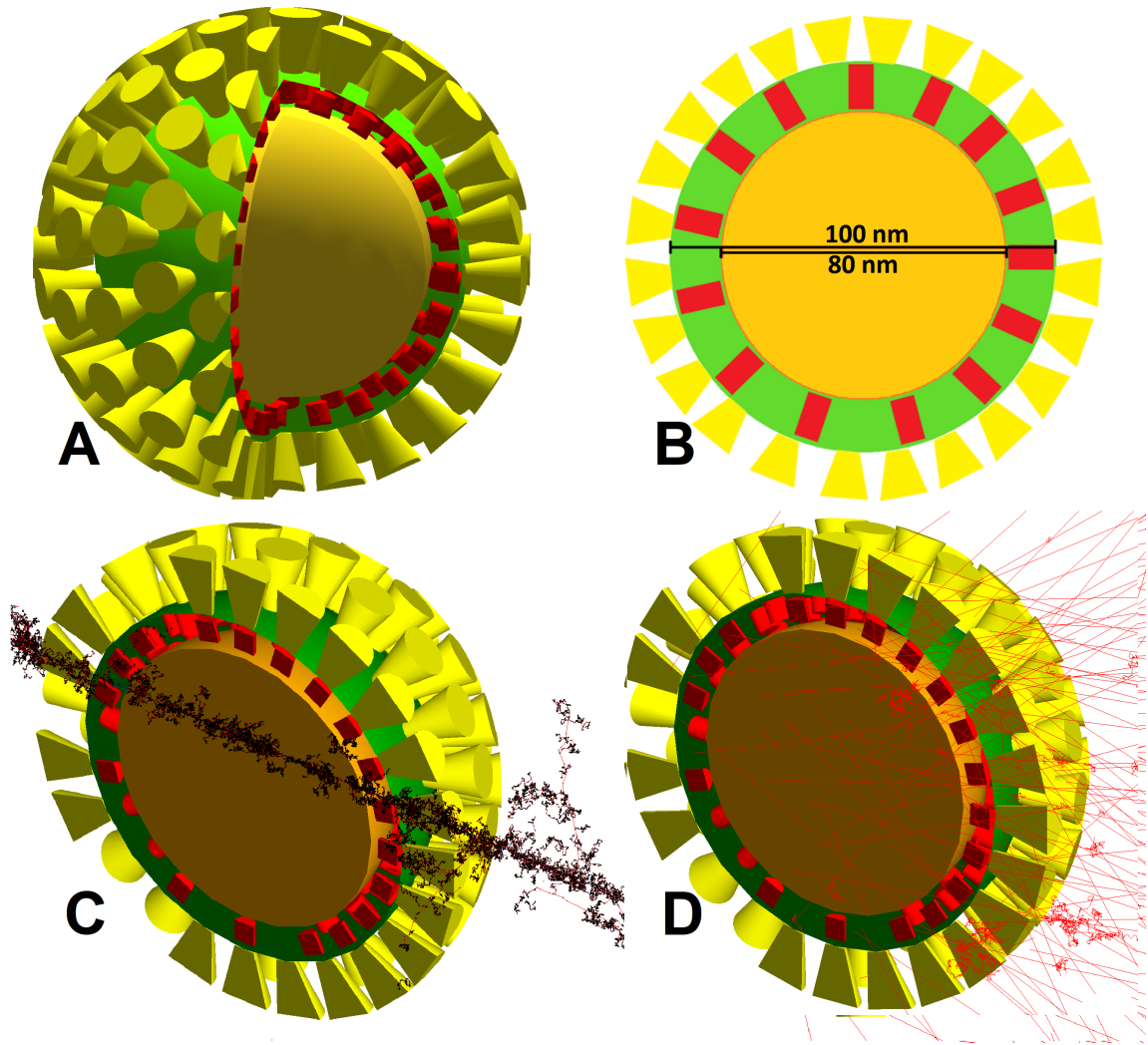

**FIG. 1.** Geometry of the virus model implemented in the simulation. Yellow cones represent the spike proteins, the membrane proteins, shown in red, are distributed within the membrane shown in green (A and B). The inner sphere represents the inside volume of the virus that usually holds

the RNA. A 450MeV/amu iron ion track is also shown crossing the virus volume with secondary electrons tracks (C) and cobalt γ-electrons are shown in (D).

The first bin of the distribution, showing the probability that no damage is induced, is highest for γ-electrons. This bin constitutes ~90% of the total damage probability, meaning that the probability of inducing at least one RNA damage is only 10%. For Fe-ions the damage probability is almost 77% of the whole spectrum, and it becomes 99.8% if only ion traversals though the nuclear envelope are considered.

The number of RNA lesions per incident particle reaches up to ~40 damage points for Fe-ions with a probability of $10^{-7}$ per eV, while the largest number observed for γ-electrons is ~9 with a probability of $10^{-7}$ per eV. Since these numbers are of very low probability and for the sake of clarity, the figure visible range was limited to 30 damage points showing the main differences between ions and γ-electrons damage spectra.

Figure 2 (right column) also shows the distribution of the number of damaged spike proteins for particles and γ-electrons, normalized per 1 eV of energy deposition, showing a higher damage yield for densely ionizing charged particles. Ideally, we would be looking for a radiation type leading to a high RNA damage yield and a small number of damaged proteins.

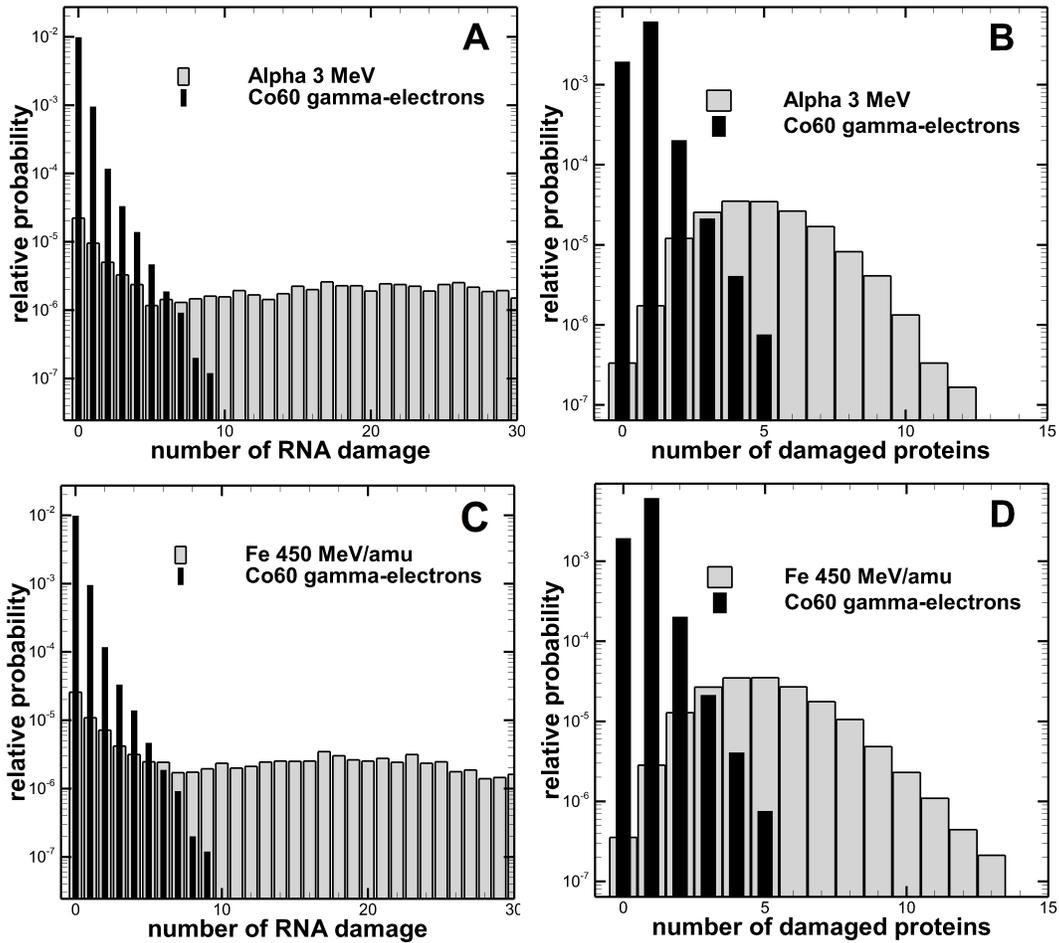

**FIG. 2.** Distributions of the number of RNA damage events, and the number of damaged spike proteins. Results are shown for alpha particles and Co60 γ–electrons (A and B), Fe-ions and Co γ–electrons (C and D), results are normalized per energy deposition of 1 eV.

However, the apparently higher number of damaged proteins is simply due to the dense ionizing track that has a high probability to damage any protein when traversing it. In Figure 2 we focus on the damage in the nucleic acid (single-stranded RNA) and the spike proteins, which is considered the main antigen target for SARS-CoV-2 vaccine.

## *Ratio of protein/RNA damage*

The average number of damaged spike proteins normalized to single RNA break is shown in Figure 3 for different incident particles according to their LET. In addition to the particles shown in Figure 2, we have included protons at 30 MeV and 80 MeV, accelerated at cyclotrons for radionuclide production or eye tumor therapy; $^{12}$C-ions at 80 MeV/amu, available in several research and medical accelerators *(51)*; and electrons, that are used for sterilization of surfaces *(52)*. The electron energy of 200 keV was selected since that electron beam studied experimentally as an alternative to γ-rays for vaccine production *(53)*. A second electron energy of 2 keV was also selected based on the results of the recent Monte Carlo study by Feng et al. *(33)* that shows that at this energy there is a maximum efficiency in virus sterilization by electrons. We found a decreasing damage ratio with increasing LET, and the minimum value, ~0.4, is obtained using α-particles and iron ions. The ratio for γ-rays and 200 keV e⁻ is ~6-7. The high ratio obtained for conventional γ-rays is due to a low average value of the RNA damage distribution.

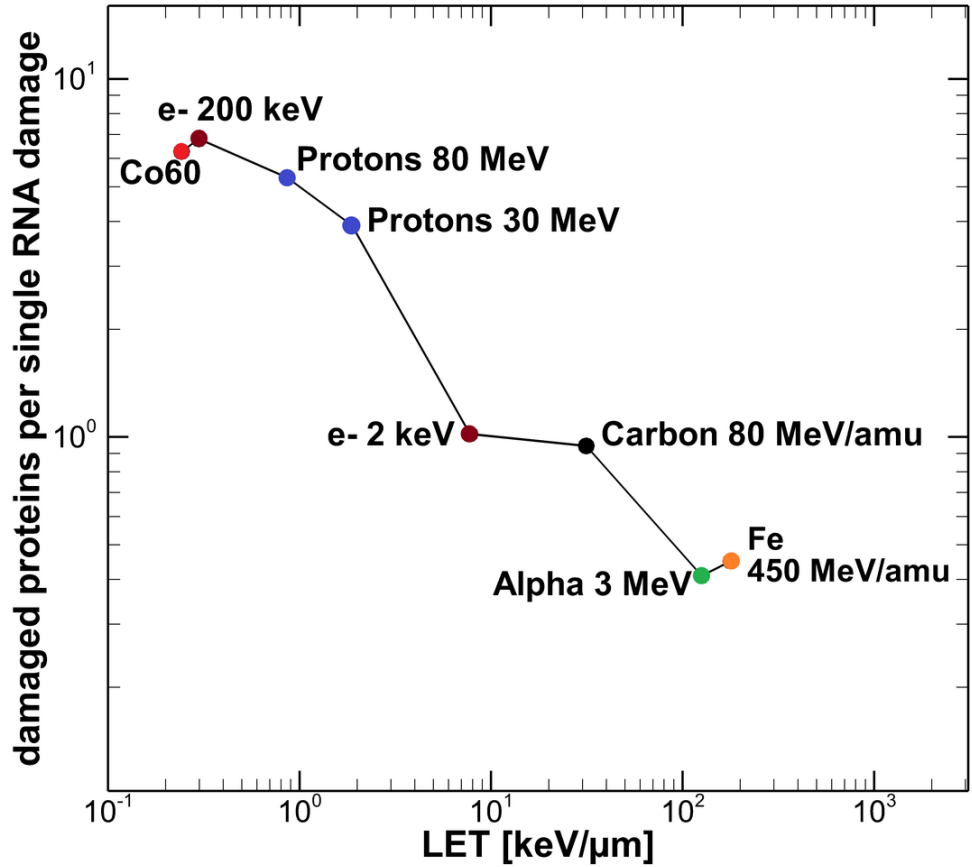

**FIG. 3.** Protein damage assessment for different particles and subsequent virus survival rates. Number of damaged spike proteins per single RNA break versus particle's LET.

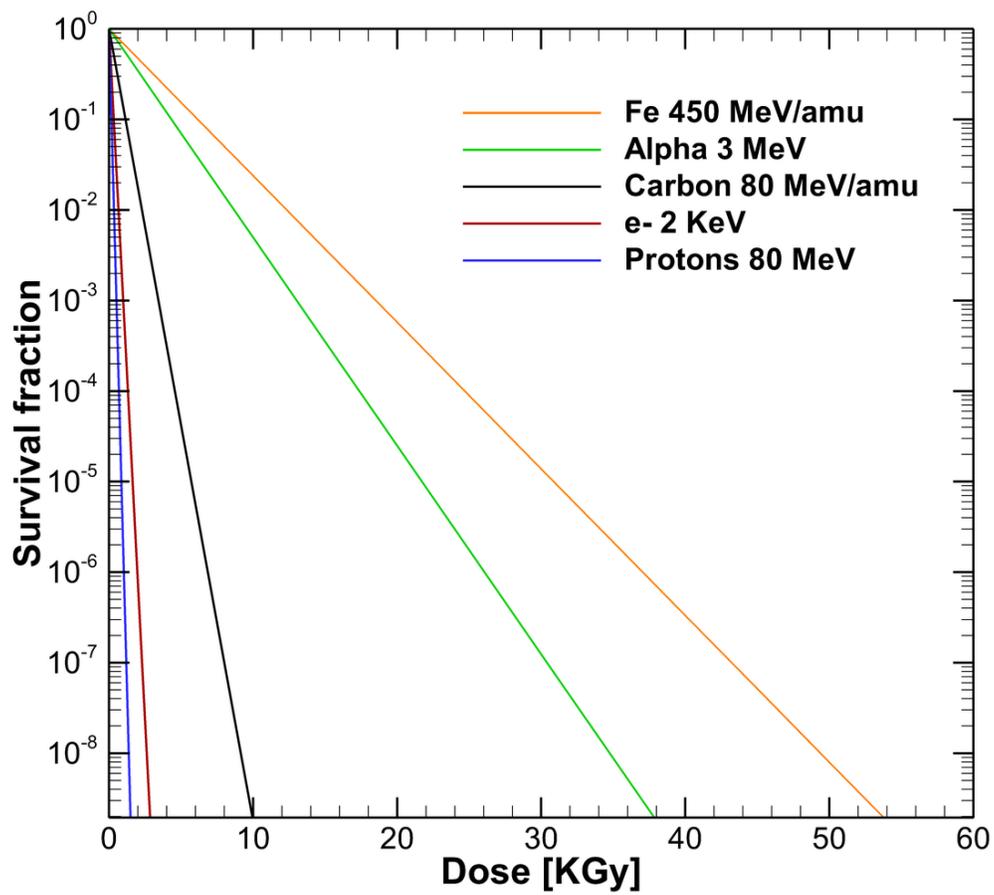

**FIG. 4.** Estimated survival curves versus irradiation dose for different particle types.

The estimated survival curves were calculated considering that one RNA single-strand break is sufficient to inactivate the virus, as it cannot be repaired and will result in an interrupted amino acids chain. Thus, the probability of inducing at least one RNA damage was used in the exponential survival model:

$$S = e^{-\sigma F}$$

where $\sigma$ is the cross-section for RNA single-strand breaks induction in the virus (calculated with the Monte Carlo simulation) and $F$ is the incident particle fluence. Exponential survival curves are essentially always found in irradiation of viruses or proteins, lacking repair capability *(13)*.

Figure 4 shows the SARS-CoV-2 survival probability as a function of the dose for sparsely and densely ionizing particles. As expected, high-LET radiation requires a much higher dose to inactivate the virus because the dose is proportional to the LET and the particle fluence must remain high enough to hit all the targets.

## *Robustness of the simulation*

As explained in the Materials and Methods section, calculations presented in Figures 2, 3 and 4 were obtained using a 10 eV threshold of the biological molecule damage. Even if this number is sound, variations in the range 5-20 eV have been used in the literature *(37,54,55)*, and it is therefore necessary to assess the robustness of our simulation to this parameter. Figure 5 shows the average number of damaged spike (A) and membrane (C) proteins normalized by single RNA break versus the energy damage threshold using values of 5 eV, 10 eV, 15 eV, and 20 eV. The ratio increases with increasing damage threshold for all radiation qualities. In fact, increasing the threshold discards all the ionization points below a certain value. Considering that each RNA ionization is a damage event in the absence of repair mechanisms (as happens in mammalian cells), increasing the threshold simply decreases the overall number of RNA damage events. Accounting for the lowest ionization binding energy in Geant4-DNA being 10.73 eV, it is in fact expected that damage ratios increase as the threshold value rises above 10 eV.

The robustness analysis should consider whether the LET dependence of the ratio (Figure 5) is affected by the modified protein/RNA ratio. Figure 5 (B and D) shows the ratio of the results obtained for the different particles in the left panel, for spike and membrane proteins respectively, formalized to the reference radiation source (γ-rays). Results for 5 eV and 10 eV thresholds are almost similar, since ionization is triggered at 10.73 eV but above this value the increase in the damage ratio becomes more significant. Notably, the ordering of the different particles is not influenced by the threshold except for electrons versus carbon ions above a threshold of 15 eV, which is seen in both parts related to spike proteins in Figure 5 (A and B).

## Discussion

Simulation of radiation damage in SARS-CoV-2 is interesting for many reasons. Here, we analyze the potential use of ionizing radiation for virus inactivation in vaccine manufacturing. As with other chemical agents, conventional inactivation with γ-rays unavoidably damages the membrane proteins, whereas ideally an inactivated virus with intact membrane antigens would elicit the most effective vaccine response. Here, we show that using densely ionizing charged particles, such as accelerated heavy ions or α-particles, the ratio of protein/RNA damage is reduced by more than an order of magnitude (Figure 3).

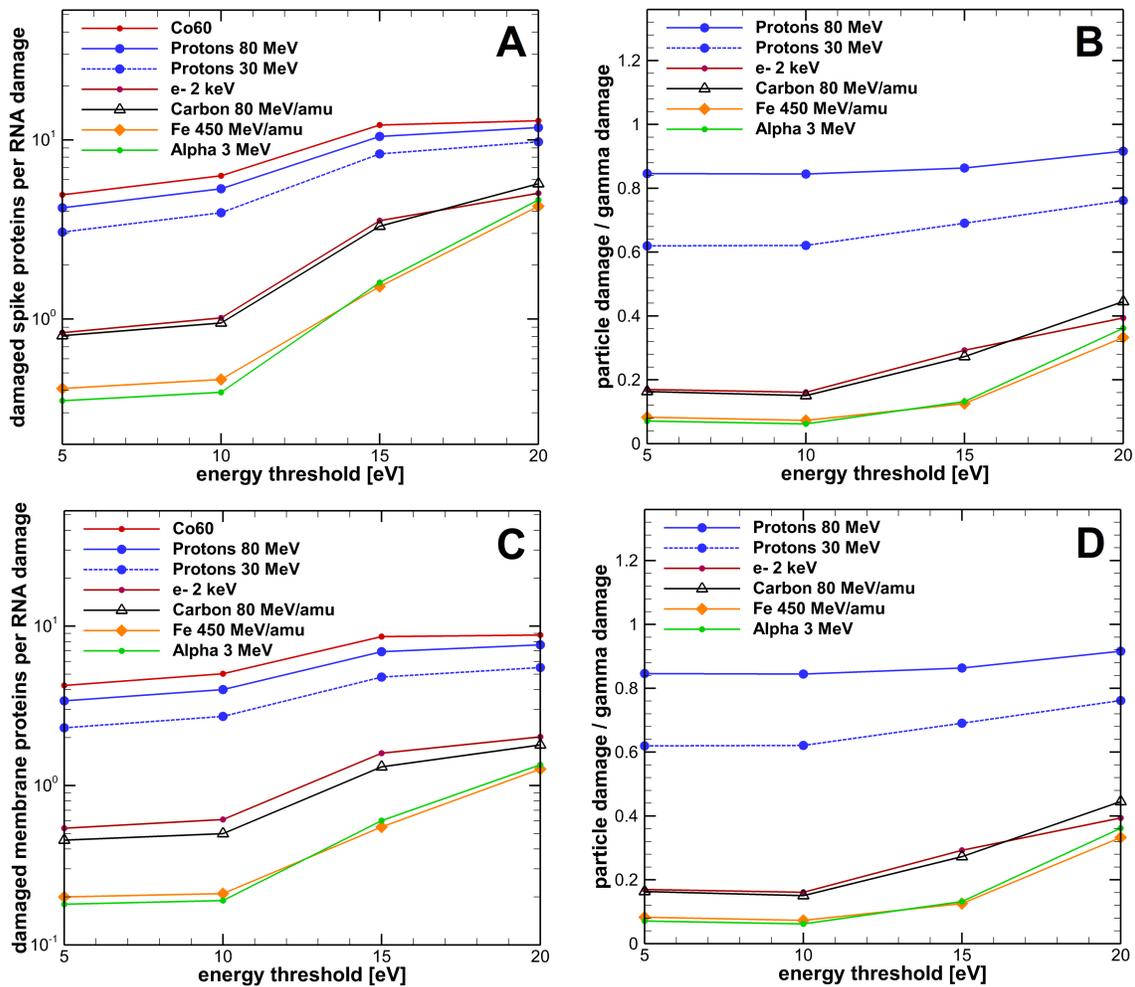

**FIG. 5.** Proteins and membrane damage comparison for different irradiation types. Number of spike (A) and membrane (C) proteins damage per RNA break calculated for different damage energy thresholds and for different incident particle types. Ratios of protein damage for the different

particles to protein damage obtained with gammas are also shown for spike proteins (B) and membrane proteins (D).

Charged particles can, therefore, represent a powerful tool for developing SARS-CoV-2 and other vaccines based on inactivated virus, which is still one of the strategies currently used for pandemics.

While Geant4-DNA is validated as an accurate Monte Carlo code for microscopic calculations, every model is affected by uncertainty, and in particular the minimum energy required to induce damage. While the minimum energy for damage in nucleic acids, characterized by strand breaks, is well known, this uncertainty is even higher for proteins, where the energy required for a conformational change, leading to a failed recognition as an antigen, could be higher. However, we have shown that our conclusions are robust in relation to the threshold energy for damage (Figure 5), meaning the advantage of densely ionizing radiation remains, or could even be bigger by an order of magnitude, with different simulation parameters.

In this study we only considered direct radiation effects on proteins and RNA, disregarding the indirect effects from free radicals. This is acceptable for high LET radiation since most of the damage comes from direct effects. However, for low LET radiation, indirect effects are an important part of the total result especially for RNA. Moreover, the free radicals damage to proteins is still unclear and related models are not available to be included in such numerical studies. Therefore, results obtained for low LET radiation might underestimate the RNA damage by a certain percentage. Probably this would further enhance the advantages of densely ionizing radiation but, while it is possible to simulate the complete RNA geometry and the chemical species diffusion using Geant4-DNA, the interactions of free radicals with specific proteins and their effects on the role of spike proteins is as yet unknown. Moreover, the virus inactivation for vaccine production generally occurs in dry ice, and this highly reduces the influence of the indirect effects.

Since an average size of 100 nm diameter was considered for our virus geometry, we expect that our results stand for the average effect of radiation on viruses with 60 nm to 140 nm diameter. sizes as presented in this work. In fact, for large dimensions we expect an increased spacing among membrane proteins and RNA molecules, resulting in a slightly lower energy deposition rates for both targets. However, the difference is not expected to affect the ratio between proteins and RNA damages since both will have lower values. In the opposite case for smaller reported dimensions, the separation between molecules will decrease and therefore we expect a slightly higher radiation effect due to the higher molecular density, so the ratio is not expected to be drastically affected since both the membrane proteins and the RNA damage rates will increase.

While the simulations clearly suggest that accelerated heavy ions can be a powerful tool for vaccine development, the practical implementation is very difficult. Energetic charged particles require large accelerators and high doses. As shown in Figure 4, inactivation doses for heavy ions are higher than for γ-rays. Survival curves similar to those in Figure 4 were measured in the past for many different viruses *(56–58)*. The increase in lethal dose is simply due to the fact that $D(Gy)$ = LET x fluence, and the fluence (in particles per $cm^2$) must be very high to hit all the viruses in the target samples, according to Poisson statistics.

The best particle for virus inactivation with minimal membrane damage depends on the LET (Figure 3) and also on the track structure, similar to biological effects in mammalian cells *(59)*. However, the choice of the best ion will be dominated by practical considerations. Large numbers of viruses have to be irradiated in closed cryovials in dry ice to keep the virus frozen *(60)*. Particles with very small penetration range (such as α-particles and low-energy electrons) are unable to penetrate the wall of a cryovial. Concerning electrons, our results support the experimental data showing that inactivation with 200 keV electron beam maintain the antigenic properties of γ−rays *(53)*, and that in principle slow (2 keV) electrons would be more efficient *(33)*, but the short range of these electrons (in water, ∼0.45 mm at 200 keV and ∼0.2 μm at 2 keV) hampers their practical usage.

High-energy heavy ions, such as Fe-ions simulated in this paper, are more suitable for vaccine production because they are both high-LET and long range, so whole cryoboxes can be irradiated in one session. High beam intensity is required to reach a kGy-level dose in a short time. In a preliminary safety test at the SIS18 synchrotron at GSI (Darmstadt, Germany), a high-intensity and -energy synchrotron *(61)*, we found that a few kGy of 1 GeV/n Fe-ions (∼28 cm range in water) can be delivered to cryoboxes containing many frozen cryovials surrounded by dry ice in a few hours, without any damage to the container and with limited and quickly reduced sample activation. These tests show that such irradiations are feasible and can represent a completely new application of large-scale particle accelerators currently in operation and under construction worldwide *(62)*.

In conclusion, our Monte Carlo simulations show that inactivation of SARS-CoV-2 can be achieved with accelerated heavy ions minimizing the damage to the epitopes in membrane proteins. Validation experiments at accelerators will be necessary to compare the capacity to induce antibody- and cell-mediated immune responses of viruses inactivated by either heavy ions or conventional methods.

While at the moment it would be impossible to have mass production in large heavy ion accelerators, cheaper technologies might be an alternative solution to produce high-LET radiation beams, e.g., commercial cyclotrons working at high-intensity, or even high-activity α-particles sources. To reach irradiation of large amounts of viral particles, it is possible to use disposable bags and continuous flow of viruses in a thin film, a method developed for virus inactivation using 200 keV electrons and that has demonstrated to be able to inactivate effectively influenza and other viruses in a reasonably short time *(63)*. If the calculation performed in this manuscript will be supported by experimental data, we expect a large efforts to develop simple methods to collect large quantities of high-LET radiation-inactivated viruses.

## Acknowledgments

The safety test on high-dose irradiation of cryovials were performed in Cave A in the frame of FAIR Phase-0 supported by the GSI Helmholtzzentrum für Schwerionenforschung in Darmstadt (Germany). We thank Dr. Uli Weber and his crew for support in those tests. We thank IN2P3/CNRS for the support to SZ and Geant4-DNA. The authors would like to thank Prof. Dr. Frederick Currell for the valuable discussions and corrections on our first draft of this manuscript.